# Comment on 'Semi-Quantum Private Comparison Based on Bell States'

You-Lin Chen[1], Yu-Chin Lu[2], Zhong-Xuan Lin[3], Tzonelih Hwang[4] (corresponding author)

*Department of Computer Science and Information Engineering, National Cheng Kung University, No. 1, University Rd., Tainan City, 70101, Taiwan, R.O.C.*

[1]p76081035@gmail.com

[2]m82617m@gmail.com

[3]p76081467@gs.ncku.edu.tw

[4]hwangtl@csie.ncku.edu.tw

**Abstract**—This study points out a semi-quantum protocol for private comparison using Bell states (SQPC) suffering from the double C-NOT attack and the malicious agent attack. The attacker can easily obtain information through these attacks. An improved protocol is proposed, which can effectively resist both of these attacks.

## 1. Introduction

In 2020, Jiang proposed a series of semi-quantum protocols for private comparison on Bell states (SQPC) [1]. Though there are two SQPC protocols proposed in this paper, we focus on the first protocol in Jiang's manuscript. The purpose of Jiang's first SQPC protocol is to let a Third Party (TP) compares the secrets of two participants, Alice and Bob, to be identical or not without leaking secrets to TP or anyone who is not the secret holder.

In this protocol, there are two participants, Alice and Bob, and a semi-honest TP. Both Alice and Bob are the classical users who can only perform the following two operations. (1) Reflection: simply reflect the receiving qubits. (2) Preparing qubits: prepare the qubits in the classical basis $\{|0\rangle, |1\rangle\}$. The definition of semi-honest TP is adapted from Yang et al. [2], which defines a semi-honest TP may misbehave on its own, but not allowed to conspire with other participants in the protocol. It is assumed that the quantum channel is ideal (i.e., non-lossy and noiseless).

However, this kind of semi-quantum protocol is usually vulnerable to the double C-NOT attack, and we point out that Jiang's SQPC suffers the same issue, which causes the information leaking. Also, we mention a malicious agent attack where one participant can obtain the other participant's secrets by simply measuring on target qubits.

This paper is organized as follows. Section 2 briefly reviews the first SQPC protocol in Jiang's manuscript. Section 3 points out the security issues in this protocol, and we propose some solutions to solve these problems. Finally, Section 4 will give a conclusion about this paper.

## 2. Review of Jiang's SQPC Protocol

*Preliminary:* There is a pre-shared key of length L, composed of $\{0,1\}$, shared between Alice and Bob, namely $K_{AB}$. Alice (Bob) generates a random binary number of length L, namely $R_A$ ($R_B$). The secret of length L holds by Alice (Bob) is denoted as $Secret_A$ ($Secret_B$). Four Bell states used in this protocol are presented in Eq. (1).

$$|\varphi^+\rangle = \frac{1}{\sqrt{2}}(|00\rangle + |11\rangle)$$
$$|\varphi^-\rangle = \frac{1}{\sqrt{2}}(|00\rangle - |11\rangle)$$
$$|\psi^+\rangle = \frac{1}{\sqrt{2}}(|01\rangle + |10\rangle)$$
$$|\psi^-\rangle = \frac{1}{\sqrt{2}}(|01\rangle - |10\rangle) \quad (1)$$

We briefly review the first SQPC protocol proposed in Jiang's manuscript. The protocol is described step-by-step in the following:

**Step 1**  TP prepares 2L EPR pairs in four Bell states mentioned in Eq. (1) randomly. Then he/she divides each EPR pair into two quantum sequences $S_A$ and $S_B$. $S_A$ comprises one particle from each EPR pair, while $S_B$ comprises another one. Both quantum sequences $S_A$ and $S_B$ will be sent to Alice and Bob, respectively.

**Step 2**  Alice (Bob) prepares a message $M_A$ ($M_B$), which is denoted by:

$$M_{A(B)} = \text{Secret}_{A(B)} \oplus R_{A(B)} \oplus K_{AB} .$$

Upon receiving $S_A$ ($S_B$) sent by TP, each of them will choose two modes called CTRL mode or SIFT mode randomly to operate. Specifically, in CTRL mode, the participant should reflect the receiving particle directly. While in SIFT mode, the participant should discard the receiving particle, then prepare a qubit according to $M_i$ ($i \in \{A, B\}$) instead. For more information, if the bit going to be encoded is 0 (1), the corresponding qubit should be prepared in $|0\rangle$ ($|1\rangle$) state. After finishing the chosen operation, the participants send back these (modified) qubits to TP.

**Step 3**  Upon receiving all the qubits sent from Alice and Bob, TP informs these participants about its receipt. Then, both participants declare about what mode (CTRL or SIFT) they chose on each particle.

**Step 4**  Depends on what mode the participants chose on each particle, TP performs a different operation.

(a) If both Alice and Bob chose CTRL mode, TP will perform a Bell measurement. In expectation, TP should get the same result as what it prepared; otherwise, there are eavesdroppers.

  (b) If any of them chose SIFT mode, TP will perform single-particle measurement on the SIFT one to get $M_A$ or $M_B$.

**Step 5**   TP checks the error rate in case (a) given in Step 4, if the value higher than threshold, this session will be aborted; otherwise, they will proceed to the next step.

**Step 6**   If the eavesdropper checking procedure succeeds, Alice and Bob publish $R_A$ and $R_B$ respectively. Then, TP computes $M_T = M_A \oplus M_B \oplus R_A \oplus R_B$ bit-by-bit. For each bit, if the result bit is '0', which means the corresponding bits in both secrets are identical, TP writes down the result, and then proceeds to next bit; otherwise, if the result bit is '1', TP finds out the secrets are different and ends the procedure immediately. If the computing procedure done and $M_T$ comprises only '0', it indicates that both secrets are identical. Finally, TP tells Alice and Bob the comparison result.

## 3. Security Issues on Jiang's First SQPC Protocol

  In a mock protocol, there are two users, namely Alice and Bob, respectively. Alice sends a number of qubits to Bob. And after Bob performed some operations on qubits, he sends those qubits, whether modified or not, back to Alice. Notice that all the qubits during the process should only be in the classical basis. A typical double C-NOT attack usually follows the template that is described as follows. The eavesdropper Eve (or a semi-honest TP) prepares a $|0\rangle$ state as the target qubit in following C-NOT attack. When the attack beginning, she performs first C-NOT on the qubits that Alice first sent to Bob, then performs second C-NOT on the returning qubits. To obtain the information, Eve just simply measures the qubits she holds, and then compares to the original state $|0\rangle$. What's more, the qubit being attacked will remain the expected state, so the information leakage will remain under covered, which is a severe security issue to any protocol.

**3.1 Double C-NOT attack on Jiang's first SQPC protocols**

Jiang's first SQPC protocol is exactly suffering from the double C-NOT attack. We take TP and Alice for example. In this example, without loss of generality, the Bell states TP preparing are all in $|\varphi^+\rangle$ state. While TP sends the particle to Alice, the attacker performs the first C-NOT attack. The states after the attack are shown in Eq. (2).

$$\frac{1}{\sqrt{2}}(|0_A 0_E 0_B\rangle + |1_A 1_E 1_B\rangle) \tag{2}$$

The subscripts of the states in Eq. (2) represent the one who holds the qubit. 'A' indicates Alice, 'B' indicates Bob, and 'E' indicates the attacker.

Then, the attacker performs another C-NOT attack while Alice is sending her qubit back to TP. Depends on the different modes Alice performed, there will be two situations occurring.

If Alice performs CTRL mode, that is, simply reflects the qubits back to TP. The states after the attack are shown in Eq. (3).

$$\frac{1}{\sqrt{2}}(|0_A 0_B\rangle + |1_A 1_B\rangle) \tag{3}$$

Both qubits of Alice and Bob remain unchanged, which means the attack still keeps silent. While the qubits that the attacker holds will remain in $|0\rangle$ state.

If Alice performs SIFT mode, that Alice discards the received qubit, both the attacker and Bob's qubit will become $|\varphi^+\rangle$ state that is shown in Eq. (4).

$$\frac{1}{\sqrt{2}}(|0_E 0_B\rangle + |1_E 1_B\rangle) \tag{4}$$

If Alice sends back a $|0\rangle$ state, after the second C-NOT attack, these states will become:

$$\frac{1}{\sqrt{2}}(|0_A 0_E 0_B\rangle + |0_A 1_E 1_B\rangle) \tag{5}$$

On the other hand, if Alice sends back a $|1\rangle$ state, after the second C-NOT attack, these states will become:

$$\frac{1}{\sqrt{2}}(|1_A 1_E 0_B\rangle + |1_A 0_E 1_B\rangle) \tag{6}$$

But, no matter what situation occurred, the attacker always has a 50% of chances to get $|1\rangle$ state as the measurement outcome. With this outcome, the attacker can ensure this qubit must operate under SIFT mode. Then, she can perform Z-basis measurement on the attacked qubit to get $M_A$. With $M_A$, the attacker can strike the known-plaintext attack and then obtain the pre-shared key $K_{AB}$ successfully while remaining in silent.

**3.2 Malicious agent attack on Jiang's first SQPC protocols**

In the previous section, we have mentioned that after Alice and Bob sent back their qubits, what operations will TP take according to their chosen modes in Step 4. As we can see, no matter who decides to perform SIFT mode, TP cannot check the correctness about this bit by Bell measurement. As the consequences, if one of the participants, for example, Bob, decides to steal Alice's secret, he can easily achieve it by simply measuring the qubits Alice sending back corresponding to the bits when he performs SIFT mode. Since the qubits produced by Alice are in the classical basis, he can prepare the same states of qubits and can send these to TP, which makes the attack not being detected, and Bob can obtain a fraction of Alice's secret.

Here we conclude how this protocol vulnerable to these attacks. Owing to the action about discarding qubits in Step 2, there will have a certain possibility that Eve can get an indicative measurement outcome in the double C-NOT attack. Both participants use the classical basis for all returned qubits, including the encoded qubits, which creates an opportunity for malicious agents can fetch a portion of the secret without detected.

**3.3 Solutions to solve the security issues on Jiang's first SQPC protocols**

Here we suggest an improved protocol to solve these security issues. In general, we change the quantum source from Bell states to single photons, alongside the new SIFT mode which uses measure-resend instead of discard-and-generate. Also, in order to check the integrity of these messages, we double the length of the encoded message, of which half will be used to fulfill this target. The detail is in the following.

First of all, as we mentioned before, the improved protocol will use single photons as the quantum source. More specifically, X-basis photons are prepared in random states $\{|+\rangle, |-\rangle\}$, the number of initial photons is 8L, where each participant will receive a half.

In Step 2, participants perform measure-resend with the classical basis in improved SIFT mode. The photons they are going to resend will be used to be $R_i$ but not encoded with $M_i$ ($i \in \{A, B\}$). This means $R_i$ increases from L to 2L.

In Step 4, for the situation that any participants chose CTRL mode on the corresponding qubits, TP performs X-basis measurement on these qubits to check if the states identical to the initial ones. If not, this session will be aborted; otherwise, they will proceed to the next step.

Here we insert an additional checking procedure. Both participants choose about half of $R_i$ and publish their position and measurement results for TP checks if there exists a malicious agent trying to strike the blocking attack on the opposite side. The blocking attack, in detail, is a malicious agent may perform X-basis measurement on all the returned qubits from the attacked side, resulting in receiving the wrong information by TP while he/she is not noticing the strikes and the loss of the integrity.

In Step 6, both participants publish $M_i$ instead of $R_i$. The remaining procedure is identical to the original.

This improved protocol, however, heavies the users' burden by adding the measurement capabilities and decreases the qubit efficiency from $\frac{1}{2}$ to $\frac{1}{4}$.

## 4. Conclusions

This paper points out a double C-NOT attack and a malicious agent attack on Jiang's first SQPC protocol. By using the double C-NOT attack, an eavesdropper Eve can steal 50% of participants' secrets while undetected. And for the malicious agent attack, the agent who plans to steal another agent's secret can actually steal a portion of secrets by simply measuring the qubits the attacked one sends back corresponding to the bits he/she performs SIFT mode. To solve these security issues, we suggest an improved protocol. By using single photons and measure-resend, although this protocol will heavy users' burden and decrease the qubit efficiency by half, it can avoid these attacks effectively and secure the information without using Bell states. Our improved protocol is also lighter than Jiang's second SQPC protocol, which is secure but the participants need to perform reorder.

## Acknowledgments

This research was partially supported by the Ministry of Science and Technology, Taiwan, R.O.C. (109-2221-E-006 -168 -)